\documentclass[12pt]{article}
\newcommand{\fr}{ \frac}

\begin{document}
\begin{center}
{\Large\bf Casimir Energy for a Wedge with Three Surfaces and for
a Pyramidal Cavity }
\end{center}
\vspace{5mm}
\begin{center}
H. Ahmedov and I. H. Duru
\end{center}

\noindent Feza G\"ursey Institute, P.O. Box 6, 81220,
\c{C}engelk\"{o}y, Istanbul, Turkey \footnote{E-mail :
hagi@gursey.gov.tr and duru@gursey.gov.tr}.

\vspace{5mm}
\begin{center}
{\bf Abstract}
\end{center}
 Casimir energy calculations for the conformally
coupled  massless scalar field for a wedge defined by three
intersecting planes and for a pyramid with four triangular
surfaces are presented. The group generated by reflections are
employed in the formulation of the required Green functions and
the wave functions.

\vspace{2cm}\noindent {\bf I. Introduction }
\\
\\
Having new geometries in hand for which we can evaluate the
Casimir energies is of interest: We learn more about the phenomena
itself, and hope that experimentalists may realize some of the
geometries to measure the effect. At this point we like to
emphasize that all experiments so far performed are of two body
ones,  a single cavity measurement has not been done. \cite{Mil}
\\
\\
For geometries with planar boundaries, if the planes are parallel
or perpendicular to each other, the method of images is easily
applicable. The parallel plates and in general rectangular prisms
 of any dimensions are of that type \cite{CUBE}. For these
geometries the groups generated by  reflections are abelian. If
the walls of the cavity are not perpendicularly intersecting, the
reflection groups are not commutative. This was the case for a
previously studied triangular region \cite{AHMED}. If the
reflections in the geometry we consider generate a non-abelian
group, one has to study the structure of this group, for it is
essential in the construction of the required Green functions and
wave functions. For example in the present study the octahedral
group provides the basic tools for the calculations.
\\
\\
 In coming  section we first calculate the Casimir energy
density for the massless scalar field  for  a wedge with three
boundary surfaces.
\\
\\
 Section III is devoted to the calculation of the Casimir energy in a pyramidal cavity
with four triangular surfaces.We get a positive result.
\\
\\
The corresponding  groups generated ( related to the Octahedral
group ) by reflections play vital role in the construction of the
Green  function and the wave functions.  The required group
theoretical details are given in the appendices.
\\
\\
{\bf II. Pyramidal wedge }
\\
\\
Consider the region in the first quadrant ( $x_1>0, \ x_2>0, \
x_3>0$ ) inside the following three planes ( Figure )
\begin{equation}\label{planes}
P_1: \ x_1=x_3, \ \ \ P_2: \ x_2=0. \ \ \ P_3: \ x_1=x_2.
\end{equation}
Reflection operators with respect to these planes are
\begin{equation}
q_1 = \left(
\begin{array}{ccc}
0 &  0& 1  \\
0 & 1 & 0 \\
1 & 0 & 0
\end{array}
\right ),
q_2 = \left(
\begin{array}{ccc}
1 &  0 & 0  \\
0 & -1 & 0 \\
0 & 0 & 1
\end{array}
\right ),
q_3 = \left(
\begin{array}{ccc}
0 & 1 & 0  \\
1 & 0 & 0 \\
0 & 0 & 1
\end{array}
\right )
\end{equation}
A group $G$ of order 48 is generated by the above reflections.
It's elements are $g_j$ and $ig_j$; $j=0, \ldots 23$. Here 24
elements $g_j$ form the Octahedral group ( see Appendix A ) and
$i=-1$ is the generator of the inversion group.
\\
\\
Consider the function
\begin{equation}\label{Green}
K(x,y) =\sum_{j=0}^{23} [ G( g_j x, y)-G( ig_j x, y)],
\end{equation}
where $G(x,y)$  is the Green function for the massless scalar
field in the Minkowski space
\begin{equation}
G(x,y)=\fr{1}{4\pi^2} \fr{1}{\mid x-y\mid^2}.
\end{equation}
Here $x$ and $y$ are  four vectors with interval $\mid x-y\mid^2 =
\mid \vec{x}-\vec{y}\mid^2-(x_0-y_0)^2$; and, $g_j$ act only on
the spatial components. To check the boundary conditions let us
consider ( with  $q_1=ig_{21}$ )
\begin{equation}\label{anty}
K(q_1x,y) =\sum_{j=0}^{23} [ G( ig_jg_{21} x, y)-G( g_jg_{21} x,
y)].
\end{equation}
Since for any $j$ we have the element  $g_k=g_jg_{21}$ in $G$,
(\ref{anty}) is equal to
\begin{equation}
K(q_1x,y) =\sum_{k=0}^{23} [ G( ig_k x, y)-G( g_k x, y)]
\end{equation}
 which implies
\begin{equation}
K(q_1 x, y)=-K(x,y).
\end{equation}
In a similar fashion one can verify the antisymmetry property for
the elements $q_2$ and $q_3$. Therefore the function $K(x,y)$
satisfies Dirichlet boundary conditions on the planes of
(\ref{planes}).
\\
\\
Note that in order the equation (  with $\eta = diag (-1, 1,1,1)$
)
\begin{equation}
  \eta^{\mu\nu}\frac{\partial^2}{\partial x^{\mu}\partial x^{\nu}}
  K(x, y) = \delta (x-y)
\end{equation}
to be satisfied by (\ref{Green}), every point in the region
between the  planes $P_1$, $P_2$ and $P_3$ must  represent
different orbits under the action of the group $G$, which is
indeed the case. In other words the region we consider is the
fundamental domain of the group $G$ ( see Appendix B ).
\\
\\
To obtain the energy momentum tensor for the  conformally coupled
massless scalar field we employ the well known coincidence limit
formula  \cite{DAVIES}
\begin{equation}
T_{\mu\nu}=\lim_{x\rightarrow y}
[\frac{2}{3}\partial^{y}_{\mu}\partial^{x}_{\nu}-\frac{1}{6}(\partial^{x}_{\mu}\partial^{x}_{\nu}+
\partial^{y}_{\mu}\partial^{y}_{\nu}) -
\frac{\eta_{\mu\nu}}{6}\eta^{\sigma\rho}\partial^{y}_{\sigma}\partial^{x}_{\rho}+
\frac{\eta_{\mu\nu}}{24}\eta^{\sigma\rho}(\partial^{x}_{\sigma}\partial^{x}_{\rho}+\partial^{y}_{\sigma}\partial^{y}_{\rho})]
K(x,y)
\end{equation}
The energy density $T(x)=T_{00}$ is given by:
\begin{equation}\label{ED}
 T(x) =\frac{1}{12\pi^2}\sum_{j=1}^{23}[ T(g_j)-T(ig_j)]
\end{equation}
 where
\begin{equation}\label{T}
T(g)=(\frac{tr(g)-1}{\mid\vec{\eta}\mid^4}-2\frac{|((1+g)\vec{\eta}|^{2}}{\mid\vec{\eta}\mid^6})
\end{equation}
 and
\begin{equation}
  \vec{\eta} = (1-g)\vec{x}
\end{equation}
with $g$ standing for $g_j$ and $ig_j$.
\\
\\
Using the invariance of $T(g)$ under the  $g\rightarrow g^{-1}$ we
have
\begin{eqnarray}\label{INV}
T(g_1) &=& T(g_3) \nonumber \\
 T(g_4) &=& T(g_6) \nonumber \\
T(g_{7}) &=& T(g_9) \nonumber \\
T(g_{10}) &=& T(g_{11}) \\
T(g_{12}) &=& T(g_{13}) \nonumber \\
T(g_{14}) &=& T(g_{15}) \nonumber \\
T(g_{16}) &=& T(g_{17}) \nonumber
\end{eqnarray}
The same is true for elements $ig_j$, with $j$ running the same
values as (\ref{INV}).
\\
\\
$\bullet$ For $g_1$ which is  rotation  by  angle $\fr{\pi}{2}$
around the $x_2$-axis we  have
\begin{equation}\label{1}
T(g_1)=-\fr{1}{(x_1^2+x_3^2)^2}
\end{equation}
\begin{equation}\label{10}
T(ig_1)=-\fr{3(x_1^2+x_3^2)+2x_2^2}{2(x_1^2+x_3^2+2x_2^2)^3}.
\end{equation}
$g_4$ and  $g_6$  are   the rotations by the same angle around the
$x_1$-axis and $x_3$-axis respectively. Therefore, $T(g_4)$,
$T(ig_4)$  and  $T(g_6)$,  $T(ig_4)$ are obtained  from  (\ref{1})
and (\ref{10}) with the cyclic replacements of coordinates
$(x_1,x_2,x_3)\rightarrow (x_3,x_1,x_2)$ and
$(x_1,x_2,x_3)\rightarrow (x_2,x_3,x_1)$ respectively.
\\
\\
$\bullet$ For the rotation $g_{12}$  by  angle $\fr{2\pi}{3}$
around the line passing trough the origin and the point $(1,-1,1)$
we have
\begin{equation}\label{2}
T(g_{12})= -\fr{3}{((x_1+x_2)^2+(x_2+x_3)^2+(x_3-x_1)^2)^2}
\end{equation}
\begin{equation}\label{20}
T(ig_{12})=-\fr{6|\vec{x}|^2+2(x_1x_3-x_1x_2-x_2x_3}{((x_1-x_2)^2+(x_2-x_3)^2+(x_3+x_1)^2)^3}.
\end{equation}
Since $g_{14}$ and $g_{16}$ are rotation matrices by the same
angle around the axis passing trough the origin and the points
$(-1,1,1)$ and $(1,1,-1)$ we conclude that $T(g_{14})$,
$T(ig_{14})$  and $T(g_{16})$, $T(g_{16})$  are given by (\ref{2})
and  (\ref{20}) with the cyclic replacements of coordinates
$(x_1,x_2,x_3)\rightarrow (x_3,x_1,x_2)$ and
$(x_1,x_2,x_3)\rightarrow (x_2,x_3,x_1)$ respectively.
\\
\\
$\bullet$ We also have
\begin{eqnarray}
T(g_{10}) = -\fr{3}{((x_1-x_2)^2+(x_2-x_3)^2+(x_3-x_1)^2)^2}, \\
T(ig_{10})= -\fr{6\mid \vec{x}\mid^2 -2(x_1x_2+x_1x_3+x_2x_3)}
{((x_1+x_2)^2+(x_1+x_3)^2+(x_2+x_3)^2)^3}.
\end{eqnarray}
\\
\\
$\bullet$ For elements satisfying the condition $g^2=1$ the second
expression in (\ref{T}) vanishes. These are the elements $g_j$,
$j=2,5,8$ and  $18,\dots, 23$.  Since $tr (ig_j)=1$  we have
$T(ig_j)=0$. Nonzero one is
\begin{equation}
 T(i)=-\frac{1}{4|\vec{x}|^4}.
\end{equation}
For  rotations $g_{20}$, $g_{21}$ and $g_2$ we get
\begin{eqnarray}\label{3}
T(g_{20})&=&  -\fr{1}{2((x_1-x_3)^2+2x_2^2)^2}, \\
T(g_{21})&=&  -\fr{1}{2((x_1+x_3)^2+2x_2^2)^2},
\end{eqnarray}
and
\begin{equation}
 T(g_2)=-\fr{1}{8(x_1^2+x_3^2)^2}.
\end{equation}
 Remaining six terms $T(g_{18})$, $T(g_{19})$, $T(g_{22})$,
$T(g_{23})$ , $T(g_5)$ and $T(g_8)$ are   obtained from the above
three equations  by the cyclic replacements of coordinates.
\\
\\
Energy density (\ref{ED}) is the given by
\begin{eqnarray}\label{final}
 T(x) = \frac{1}{12\pi^2} [ T(g_{10})- T(ig_{10})-T(i)] + \ \ \ \ \
 \ \ \ \ \ \ \ \ \ \ \ \ \ \ \ \ \ \ \ \ \ \ \ \ \ \ \ \ \ \ \ \ \ \ \ \ \\
+\frac{1}{12\pi^2}[ \frac{17}{8}T(g_1) -2T(ig_1)+2T(g_{12})  -
2T(ig_{12})+T(g_{20}) +T(g_{21})+ c.p. ]\nonumber
\end{eqnarray}
where $c.p.$ stands for cyclic permutations of coordinates. The
system we consider is the intersection region of three wedges
$(P_1, P_2)$, $(P_2, P_3)$ and $(P_1,P_3)$. For  $x_3\gg 1$ and
$\sqrt{x_1^2+x_2^2}\ll x_3$ our result should reduce to the wedge
problem $(P_2,P_3)$. Recall, that energy density in the wedge
between two inclined planes is \cite{WEDGE}
\begin{equation}\label{w0}
T_W = -\fr{1}{1440\pi^2r^4}(\fr{\pi^4}{\alpha^4}-1),
\end{equation}
where $\alpha$ is the angle between two planes and $r$ is the
minimal distance to the  axis which is intersection of two planes.
For the system $(P_2,P_3)$ we have $\alpha =\fr{\pi}{4}$ and
$r^2=x_1^2+x^2_2$, that is (\ref{w0}) takes the form:
\begin{equation}\label{w}
T_{P_2P_3}= -\fr{255}{1440\pi^2(x_1^2+x^2_2)^2}.
\end{equation}
Coming to our density (\ref{final}) for $x_3\gg 1$ and
$\sqrt{x_1^2+x_2^2}\ll x_3$ all terms  except the one with
$T(g_6)$ are  negligibly small; thus, we have
\begin{equation}
T(x)\simeq -\frac{17}{12\pi^2 \cdot
8}T(g_6)=-\fr{17}{96\pi^2(x_1^2+x_2^2)^2}
\end{equation}
which is same as (\ref{w}). In a similar fashion, terms
$T(g_{20})$ and $T(g_{10})$ correspond in the suitable limits to
the wedge problems $(P_1P_2)$ and $(P_1P_3)$.
\\
\\
{\bf III. Pyramid}
\\
\\
We add  to the planes $P_1, P_2, P_3$ the fourth one $P_4: x_3=a$.
Reflection with respect to the $P_4$ plane is given by
\begin{equation}
q_4 \left(
\begin{array}{c}
x_1  \\
x_2 \\
x_3
\end{array}
\right ) = \left(
\begin{array}{c}
x_1 \\
x_2 \\
2-x_3
\end{array}
\right )
\end{equation}
The group  generated by $q_j$, $j=1,2,3,4$ is the semidirect
product of the group $G$ defined in the previous section and  the
translation group $Z^3$. The Green function vanishing on the
planes $P_a$, $=1,\dots ,4$ is given by
\begin{equation}
K(x,y) = \sum_{n,m,k=-\infty}^\infty \sum_{j=1}^{23} [ G( g x
+\xi, y)-G( g x +\xi, y)]
\end{equation}
where
\begin{equation}
\xi = \left(
\begin{array}{c}
0 \\
2na  \\
2ma \\
2ka
\end{array}
\right ).
\end{equation}
The energy momentum density is
\begin{equation}
 T(x) =\frac{1}{6\pi^2}\sum_{n,m,k=-\infty}^\infty  \sum_{j=1}^{23} [
 T(g_j)-T(ig_j)]
 \end{equation}
where $T(g)$ is given by (\ref{1}) with the  replacement
$\vec{\eta} \rightarrow \vec{\eta} + \vec{\xi}$, where $\vec{\xi}$
is the spatial part of the four dimensional vector $\xi$. We
calculate explicitly  the total vacuum energy of the pyramid.
 For the geometry in hand it is reasonable to use another representation for the Green function
related to the wave function and the spectra of the quantum
mechanical system inside the pyramid. The wave  function which
vanish on the planes $P_1$, $P_2$ and $P_3$ can be obtained in a
similar fashion as the Green function:
\begin{equation}\label{wave0}
\Psi (\vec{x})= \Omega \sum_{g\in G} \sum_{j=1}^{23}[
e^{i(\vec{p},g\vec{x})}-e^{i(\vec{p},ig\vec{x})}]
\end{equation}
or
\begin{equation}\label{wave}
\Psi_{\vec{p}} (\vec{x})= -8i\Omega [\sin p_1x_1\sin p_2x_2\sin p_3x_3-
\sin p_1x_1\sin p_2x_3\sin p_3x_2 + \ c. p. ]
\end{equation}
where $\Omega$ is the normalization.

 The condition $\Psi_{\vec{p}} (\vec{x})\mid_{P_4}=0$
implies that the components  $p_j$ are  proportional to the
nonzero positive integers
\begin{equation}\label{SP}
p_1=\frac{\pi}{a} n, \ \ p_2=\frac{\pi}{a} m, \ \
p_3=\frac{\pi}{a} k.
\end{equation}
The properties
\begin{equation}
\Psi_{g_j\vec{p}} (\vec{x})=\Psi_{\vec{p}} (\vec{x}) \ \ \
\Psi_{i\vec{p}} (\vec{x})=\Psi_{\vec{p}} (\vec{x})
\end{equation}
imply that the spectrum  takes its values in the quotient space
\begin{equation}
Z^3/G = \{ \vec{n}\in Z^3: \ k\geq n\geq m \geq 0\}
\end{equation}
which is the discrete analogue of the pyramidal region considered
in the previous section.  The wave function (\ref{wave}) vanishes
on the boundary $B$ of $A=Z^3/G$. Boundary of $A$ is the union of
three  regions : $k\geq n\geq m=0$, $k\geq n=m\geq  0$ and $k=
n\geq m\geq  0$. We have to drop these values from the spectra.
Physical spectra is given by (\ref{SP}) with $k> n> m
> 0$ or $\vec{n}\in A/B$.
\\
\\
The Green function can be written as
\begin{equation}
G(x,y)=\sum_{k=3}^\infty\sum_{n=2}^{k-1}\sum_{m=1}^{n-1}\fr{e^{i\pi\mid
\vec{p}\mid (x_0-y_0)}} {2\mid\vec{p}\mid}\Psi_{\vec{p}} (\vec{x})
\Psi_{\vec{p}} (\vec{y})
\end{equation}
which implies
\begin{equation}
 T(x)=\fr{\pi}{2a}\sum_{k=3}^\infty\sum_{n=2}^{k-1}\sum_{m=1}^{n-1}
\sqrt{n^2+m^2+k^2}\mid \Psi_{\vec{p}} (\vec{x})\mid^2.
\end{equation}
After integration $\int_0^a dx_3 \int_0^{x_3} dx_1 \int_0^{x_1}
dx_2$ we have ( with $\vec{n}=(n,m,k)$ )
 \begin{equation}
E=\frac{\pi}{2a}\sum_{\vec{n}\in A/B}|\vec{n}|=
\frac{\pi}{96a}\sum_{\vec{n}\in A/B}\sum_{g\in G}|g\vec{n}|=
\frac{\pi}{96a} \sum_{\vec{n}\in \bigcup g(A/B)}|\vec{n}|
\nonumber
\end{equation}
 Since $A$ is the quotient space $Z^3/G$ we have
\begin{equation}
 \bigcup_{g\in G} g(A/B)= Z^3/C, \ \ \ C = \bigcup_{g\in G} gB
\end{equation}
which implies
\begin{equation}
 E=\frac{\pi}{96a}( \sum_{\vec{n}\in Z^3}|\vec{n}| - \sum_{\vec{n}\in
 C}|\vec{n}|).
 \end{equation}
 $C$ is the union of nine planes: $m=0$, $n=m$, $n=k$ and other six
planes are obtained by cyclic permutations of $n, m$ and $k$ :
\begin{equation}
 \sum_{\vec{n}\in C}|\vec{n}|=3\sum_{n,m\in Z}\sqrt{n^2+m^2}+6\sum_{n,m\in Z}\sqrt{2n^2+m^2}
 \end{equation}
The Casimir energy is
\begin{equation}\label{CASIMIR}
E= \frac{1}{6}E _1-\frac{1}{2}E_2-\frac{6+4\sqrt{2}}{16}E_3 \simeq
\frac{0,069}{a}
\end{equation}
 Here $E_1$, $E_2$ and  $E_3$  are the Casimir energies for the cube with sides
 $a$,  for the  rectangle with sides $a$,
 $\frac{a}{\sqrt{2}}$ and  for the one dimensional system of length
 $a$ ( see \cite{CUBE} and references therein ) :
\begin{eqnarray}
E_1&=&\frac{\pi}{2a}\sum_{n,m,k=1}^\infty \sqrt{n^2+m^2+k^2}\simeq   -\frac{0,015}{a} \\
E_2&=&\frac{\pi}{2a}\sum_{n,m=1}^\infty \sqrt{n^2+2m^2} \simeq
\frac{0,045}{a}     \\
E_3&=&\frac{\pi}{2a}\sum_{n=1}^\infty n  \simeq -\frac{0,131}{a}
\end{eqnarray}
The positive result of (\ref{CASIMIR}) is about the same magnitude
as the other well known positive Casimir energy example of the
spherical cavity with radius $a$ \cite{BALL}
\begin{equation}
E_{ball} \simeq \frac{0,045}{a}.
\end{equation}
For nanometer size, that is for $a=10^{-7}$ cm, the energy
(\ref{CASIMIR}) is ( in $\hbar=c=1$ unit, $1 eV\cong 0,5 \otimes
10^5$ cm$^{-1}$ ) $E\simeq 35$ eV which is of considerable
magnitude.
\\
\\
{\bf Acknowledgment}: The authors thank Turkish Academy of Science
( TUBA ) for its support, and to D. Duru for the figure.
\\
\\
\setcounter{equation}{0}
\def\theequation{A.\arabic{equation}}

\vspace{1cm} \noindent {\bf Appendix A}
\\
\\
Octahedral group $O$ is the group of transformations which
transforms cube into itself. The order of this group is 24. We
denote the identity element by $g_0$. $g_1$, $g_2$ and $g_3$ are
rotations on $\fr{\pi}{2}$, $\pi$ and $\fr{3\pi}{2}$ around
y-axis:
\begin{equation}
g_1 = \left(
\begin{array}{ccc}
0 &  0 & -1  \\
0 & 1 & 0 \\
1 & 0 & 0
\end{array}
\right ),
g_2 = \left(
\begin{array}{ccc}
-1 & 0 & 0 \\
0 & 1 & 0  \\
0 & 0  & -1
\end{array}
\right ), \\
g_3 = \left(
\begin{array}{ccc}
0 & 0 & 1  \\
0  & 1 & 0 \\
-1 & 0 & 0
\end{array}
\right )
\end{equation}
$g_4$, $g_5$ and $g_6$ are rotations by $\fr{\pi}{2}$, $\pi$ and
$\fr{3\pi}{2}$ around x-axis:
\begin{equation}
g_4 = \left(
\begin{array}{ccc}
1 &  0 & 0  \\
0 & 0 & -1 \\
0 & 1 & 0
\end{array}
\right ),
g_5 = \left(
\begin{array}{ccc}
1 &  0 & 0  \\
0 & -1 & 0 \\
0 & 0 & -1
\end{array}
\right ), \\
g_6 = \left(
\begin{array}{ccc}
1 &  0 & 0  \\
0 &  0 & 1 \\
0 & -1 & 0
\end{array}
\right )
\end{equation}
$g_7$, $g_8$ and $g_9$ are rotations by $\fr{\pi}{2}$, $\pi$ and
$\fr{3\pi}{2}$ around z-axis:
\begin{equation}
g_7 = \left(
\begin{array}{ccc}
0 &  -1 & 0  \\
1 &  0 & 0 \\
0 & 0 & 1
\end{array}
\right ),
g_8 = \left(
\begin{array}{ccc}
-1 &  0 & 0  \\
0 & -1 & 0 \\
0 & 0 & 1
\end{array}
\right ), \\
g_9 = \left(
\begin{array}{ccc}
0 &  1 & 0  \\
-1 &  0 & 0 \\
0 & 0 & 1
\end{array}
\right )
\end{equation}
$g_{10}$ and  $g_{11}$  are rotations by $\fr{2\pi}{3}$ and
$\fr{4\pi}{3}$ around the axis passing trough the origin and the
point $(1,1,1)$:
\begin{equation}
g_{10} = \left(
\begin{array}{ccc}
0 &  0 & 1  \\
1 &  0 & 0 \\
0 & 1 & 0
\end{array}
\right ),
g_{11} = \left(
\begin{array}{ccc}
0 & 1 & 0  \\
0 & 0 & 1 \\
1 & 0 & 0
\end{array}
\right ),
\end{equation}
$g_{12}$ and  $g_{14}$  are rotations by $\fr{2\pi}{3}$ and
$\fr{4\pi}{3}$ around the axis passing trough the origin and the
point $(1,-1,1)$:
\begin{equation}
g_{12} = \left(
\begin{array}{ccc}
0 &  0 & 1  \\
-1 &  0 & 0 \\
0 & -1 & 0
\end{array}
\right ),
g_{13} = \left(
\begin{array}{ccc}
0 & -1 & 0  \\
0 &  0 & -1 \\
1 & 0 & 0
\end{array}
\right )
\end{equation}
$g_{14}$ and  $g_{15}$  are rotations by $\fr{2\pi}{3}$ and
$\fr{4\pi}{3}$ around the axis passing trough the origin and the
point $(-1,1,1)$:
\begin{equation}
g_{14} = \left(
\begin{array}{ccc}
0 &  0 & -1  \\
-1 &  0 & 0 \\
0 & 1 & 0
\end{array}
\right ), \\
g_{15} = \left(
\begin{array}{ccc}
0 &  -1 & 0  \\
0 &  0 & 1 \\
-1 & 0 & 0
\end{array}
\right )
\end{equation}
$g_{16}$ and  $g_{17}$  are rotations by $\fr{2\pi}{3}$ and
$\fr{4\pi}{3}$ around the axis passing trough the origin and the
point $(1,1,-1)$:
\begin{equation}
g_{16}= \left(
\begin{array}{ccc}
0 &  0 & -1  \\
1 &  0 & 0 \\
0 & -1 & 0
\end{array}
\right ),
g_{17} = \left(
\begin{array}{ccc}
0 &  1 & 0  \\
0 &  0 & -1 \\
-1 & 0 & 0
\end{array}
\right ).
\end{equation}
$g_{18}$ and  $g_{19}$  are rotations by $\pi$ around the axis
passing trough the origin and the points $(1,1,0)$ and $(1,-1,0)$
respectively:
\begin{equation}
g_{18} = \left(
\begin{array}{ccc}
0 & 1 & 0  \\
1 &  0 & 0 \\
0 & 0 & -1
\end{array}
\right ),
g_{19} = \left(
\begin{array}{ccc}
0 & -1 & 0  \\
-1 &  0 & 0 \\
0 & 0 & -1
\end{array}
\right )
\end{equation}
$g_{20}$ and  $g_{21}$  are rotations by $\pi$ around the axis
passing trough the origin and the points $(1,0,1)$ and $(1,0,-1)$
respectively:
\begin{equation}
g_{20} = \left(
\begin{array}{ccc}
0 &  0 & 1  \\
0 & -1 & 0 \\
1 & 0 & 0
\end{array}
\right ), \\
g_{21} = \left(
\begin{array}{ccc}
0 &  0 & -1  \\
0 & -1 & 0 \\
-1 & 0 & 0
\end{array}
\right )
\end{equation}
$g_{22}$ and  $g_{23}$  are rotations by $\pi$ around the axis
passing trough the origin and the points $(0,1,1)$ and $(0,1,-1)$
respectively:
\begin{equation}
g_{22} = \left(
\begin{array}{ccc}
-1 &  0 & 0  \\
0 &  0 & 1 \\
0 & 1 & 0
\end{array}
\right ),
g_{23} = \left(
\begin{array}{ccc}
-1&  0 & 0  \\
0 & 0 & -1 \\
0 & -1 & 0
\end{array}
\right ).
\end{equation}
The tetrahedral group $T$ is the subgroup of $O$ of order $12$
with elements $g_0. g_2, g_5, g_8$ and $g_{10}, \ldots g_{17}$.
For more details we refer to  \cite{Jans}.

\setcounter{equation}{0}
\def\theequation{B.\arabic{equation}}

\vspace{1cm} \noindent {\bf Appendix B}
\\
\\
Let $G$ be a point group acting in the Euclidean  space  $R^3$. A
closed subset $S$ of $R^3$ is called a fundamental domain of $G$
 if $R^3$ is the union of conjugates of $S$, i.e.,
\begin{equation}
 R^3=\bigcup_{g\in G} gS
\end{equation}
and the intersection of any two conjugates has no interior.
\\
\\
The fundamental domain of the group generated by the reflections
$ig_2$, $ig_5$ and $ig_8$ with respect to $y=0$, $x=0$ and $z=0$
planes is the first quadrant in $R^3$. This is group of order $8$
and divide $R^3$ into $8$ equal parts. If one adds to this group
the element  $g_{10}$ we arrive at the group of order $24$ which
is the direct product of the tetrahedral group $T$ and inversion
one $I$ generated by $i$. Rotation $g_{10}$ is three fold
rotation. It divides the first quadrant into three equal parts.
Therefore the fundamental domain for $T\times I$ is the region in
the first quadrant between three planes $P_2$, $P_3$ and $P_5=
\{x_2=x_3\}$. In $T\times I$ there is reflection operator $ig_2$
with respect to $P_2$ plane. The Green function constructed from
the group $T\times I$ will vanish on $P_2$. For $P_3$ and $P_5$
there is no reflection operators. $g_{10}$ rotate $P_3$ into
$P_5$. If we add to $T\times C$ the reflection operator with
respect to $P_3$ plane we arrive at the group $O\times I$ which is
of order $48$. The corresponding fundamental domain can be
obtained from that of $T\times C$ by dividing it into two equal
parts, that is the region between three planes $P_2$, $P_3$ and
$P_1$. There are reflections with respect to these planes. The
Green function will vanish on these planes.

\end{document}